\documentclass[a4paper,12pt]{article}
\usepackage[a4paper,text={16.8cm,22.4cm}]{geometry}
\usepackage{amsmath,amssymb,bm,psfrag,graphicx}
\usepackage[normalem]{ulem}
\usepackage{color}

\allowdisplaybreaks 
\renewcommand{\arraystretch}{1.2}

\newcommand{\pTveto}{p_T^{\hspace{-0.3mm}\rm veto}\hspace{-0.3mm}}
\newcommand{\pLveto}{p_L^{\hspace{-0.3mm}\rm veto}\hspace{-0.3mm}}
\newcommand{\pRveto}{p_R^{\hspace{-0.3mm}\rm veto}\hspace{-0.3mm}}
\newcommand{\dv}{d_2^{\rm veto}\hspace{-0.5mm}}
\newcommand{\ff}{f\hspace{-1.6mm}f}
\newcommand{\II}{I\hspace{-1mm}I}

\begin{document}

\begin{titlepage}

\begin{flushright}
MZ-TH/12-18\\
May 16, 2012
\end{flushright}

\vspace{0.6cm}
\begin{center}
\Large\bf
Factorization and NNLL Resummation for\\ 
Higgs Production with a Jet Veto
\end{center}

\vspace{0.3cm}
\begin{center}
Thomas Becher$^a$ and Matthias Neubert$^b$\\
\vspace{0.4cm}
{\sl 
${}^a$\,Albert Einstein Center for Fundamental Physics\\
Institut f\"ur Theoretische Physik, Universit\"at Bern\\
Sidlerstrasse 5, CH--3012 Bern, Switzerland\\[0.4cm]
${}^b$\,Institut f\"ur Physik (THEP), 
Johannes Gutenberg-Universit\"at\\ 
D--55099 Mainz, Germany}
\end{center}

\vspace{0.5cm}
\begin{abstract}
\vspace{0.2cm}
\noindent 
Using methods of effective field theory, we derive the first all-order factorization theorem for the Higgs-boson production cross section with a jet veto, imposed by means of a standard sequential recombination jet algorithm. Like in the case of small-$q_T$ resummation in Drell-Yan and Higgs production, the factorization is affected by a collinear anomaly. Our analysis provides the basis for a systematic resummation of large logarithms $\ln(m_H/\pTveto)$ beyond leading-logarithmic order. Specifically, we present predictions for the resummed jet-veto cross section and efficiency at next-to-next-to-leading logarithmic order. Our results have important implications for Higgs-boson searches at the LHC, where a jet veto is required to suppress background events.
\end{abstract}
\vfil

\end{titlepage}

\section{Introduction}

While the strong interactions generally complicate physics at hadron colliders, the Higgs-boson production cross section in gluon-gluon fusion is one of the few places where QCD is trying to help: radiative corrections substantially increase the production rate at both next-to-leading (NLO) \cite{Dawson:1990zj,Djouadi:1991tka} and next-to-next-to leading order (NNLO) in perturbation theory \cite{Harlander:2002wh,Anastasiou:2002yz,Ravindran:2003um}. However, to interpret results from Higgs searches and to study Higgs properties, it is important that the higher-order corrections are under control. In \cite{Ahrens:2008qu,Ahrens:2008nc} we have analyzed the total production cross section and argued that the largest effects arise in the virtual corrections to the $gg\to H$ amplitude, which are given by the scalar gluon form factor evaluated at time-like momentum transfer. These corrections can be resummed using renormalization-group (RG) methods.

\begin{figure}
\begin{center}
\includegraphics[width=\textwidth]{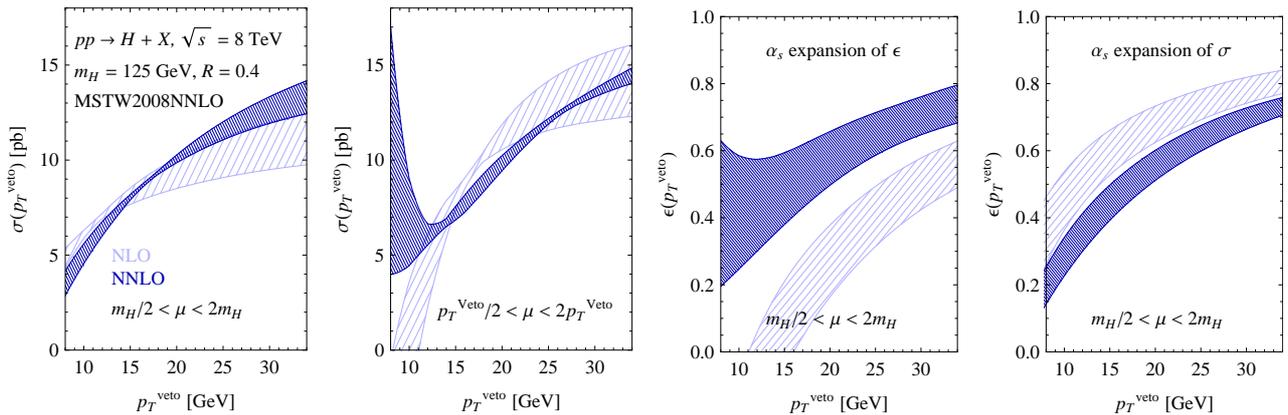}
\caption{\label{fig:resFixed} 
The jet-veto cross section $\sigma(\pTveto)$ and efficiency $\epsilon(\pTveto)$ for Higgs production at the LHC, at NLO (shaded light bands) and NNLO (dark bands) in fixed-order perturbation theory. The first two panels compare the scale choices $\mu\sim m_H$ and $\mu\sim\pTveto$. The right panels show predictions for the veto efficiency, defined either as a perturbative series in $\alpha_s$ (third plot), or as the ratio of the $n^{\rm th}$-order vetoed cross section to the $n^{\rm th}$-order total cross section (last plot).}
\end{center}
\end{figure}

The total cross section is not directly relevant for Higgs-boson searches, since these impose a jet veto $p_T^{\rm jet}<\pTveto$ to enhance the signal-to-background ratio. In particular, the jet veto is essential to reduce the background to the $pp\to H\to W^+ W^-$ process arising from top-quark production with subsequent $t\to b\,l^+\nu$ decay. To get sufficient background reduction, the experimental searches reject all events containing jets with transverse momenta larger than $\pTveto\approx 20$\,--\,30\;GeV. It was observed that the $K$-factor for Higgs production gets reduced when such a cut is imposed \cite{Catani:2001cr,Anastasiou:2004xq,Anastasiou:2007mz}. Keeping the parton distribution functions (PDFs) and the value of $\alpha_s$ fixed, the total cross section increases by more than 100\% when going from LO to NLO, and by another 35\% when going from NLO to NNLO. The corrections to the jet-veto cross section are however much smaller. This is evident from the first two plots in Figure~\ref{fig:resFixed}, which show fixed-order results for the jet-veto cross section for Higgs-boson production at the LHC, obtained using different scale choices. We employ a standard jet-clustering algorithm with $R=0.4$ (see below) and use MSTW2008NNLO PDFs \cite{Martin:2009iq}. For example, at $\pTveto=20$\,GeV the difference between the NLO and NNLO cross sections is only about 10\%. Note that in the presence of the jet veto the cross section involves the two disparate scales, $\pTveto$ and $m_H$, and it is not clear {\it a priori\/} which value of the factorization and renormalization scales one should choose (we set $\mu=\mu_f=\mu_r$ throughout). Indeed, the results look rather different depending on whether one takes $\mu=m_H$ (first plot) or $\mu=\pTveto$ (second plot) as the default value. It thus appears somewhat suspicious that the NNLO results for $\pTveto>15$\,GeV happen to be rather similar in the two cases. In \cite{Ahrens:2008nc} we have argued that the reduction of the $K$-factor observed in the presence of a jet veto arises from an accidental cancellation of the large positive corrections to the total cross section with large negative corrections from Sudakov logarithms $\alpha_s^n\ln^{2n}(m_H/\pTveto)$, which arise because the jet veto enhances the contributions from emissions collinear to the beams. 

The presence of large Sudakov logarithms becomes obvious when one considers the jet-veto efficiency $\epsilon(\pTveto)=\sigma(\pTveto)/\sigma_{\rm tot}$, defined as the ratio of the vetoed to the total cross section. In this quantity the common large virtual corrections to the $gg\to H$ amplitude drop out, leaving only the large Sudakov logarithms behind. The third panel in Figure~\ref{fig:resFixed} shows that the veto efficiency indeed receives very large higher-order corrections. Even at NNLO the remaining scale variation is rather significant. In the figure we choose a high default value for the factorization scale. Choosing $\mu\sim\pTveto$ would lead to a complete breakdown of the fixed-order expansion for the efficiency. In addition to the large scale uncertainty the result suffers from a significant scheme ambiguity. The last plot in the figure shows the predictions for the efficiency obtained when one expands the two cross sections $\sigma(\pTveto)$ and $\sigma_{\rm tot}$ to NLO or NNLO, but then takes their ratio without performing a further expansion in powers of $\alpha_s$. Comparing the last two panels, we observe that the scheme dependence remains uncomfortably large even at NNLO.

The leading-logarithmic (LL) corrections to the jet-veto cross section were studied using parton showers \cite{Anastasiou:2008ik,Anastasiou:2009bt}, but for a long time no systematic resummation of higher-order logarithmic terms was available. To improve the accuracy of the numerical predictions, the parton-shower results were re-weighted to the Higgs-boson $q_T$ spectrum obtained at next-to-next-to-leading logarithmic (NNLL) order, as implemented in the code {\sc HqT} \cite{deFlorian:2011xf}. As an alternative, it was suggested to use the event-shape variable beam thrust instead of a jet veto to discriminate against the background from top decays \cite{Berger:2010xi}. The NNLL resummed results for this variable confirmed the picture that there are large corrections from collinear emissions. It was found that a fixed-order computation is not reliable and that the scale variation underestimates the perturbative uncertainties. While beam thrust is theoretically simpler than the jet veto, it would be more difficult to use in experimental analyses, since this observable is sensitive to underlying-event and pile-up effects. Based on the results for beam thrust, an alternative way to estimate the scale uncertainties for Higgs searches using jet bins was proposed in \cite{Stewart:2011cf}. 

Very recently, however, it was pointed out that the resummation for the jet-veto cross section defined with a standard sequential jet algorithm is indeed feasible, and numerical results for the cross section at NLL order were presented \cite{Banfi:2012yh}. These results were obtained using the {\sc Caesar} code \cite{Banfi:2004yd}, which performs automated NLL resummations for observables that are subject to certain conditions, such as recursive infrared safety and globalness. In this paper, we study the resummation for the cross section in the presence of a jet veto in Soft-Collinear Effective Theory (SCET) \cite{Bauer:2000yr,Bauer:2001yt,Bauer:2002nz,Beneke:2002ph}. In Section~\ref{sec:factorization}, we propose an all-order factorization theorem, according to which the cross section factorizes into a product of a hard function, a soft function, and two beam-jet functions. This simple factorization into a product of functions (rather than a convolution) arises because, at leading order in an expansion in powers of $\pTveto/m_H$, the $k_T$-type jet algorithms do not cluster soft and collinear radiation inside the same jet. As for small-$q_T$ resummation, the naive factorization is affected by a collinear anomaly \cite{Becher:2010tm}, which induces a dependence on the Higgs-boson mass in the product of beam-jet and soft functions. We derive the all-order form of this anomaly and present simple analytic formulae for the resummed cross section. In Section~\ref{sec:oneloop}, we then give all ingredients required for carrying out the resummation at NNLL order, including a certain two-loop coefficient capturing the dependence of the resummed cross section on the jet algorithm. We extract this coefficient from partial results at NNLL order presented in the appendix of \cite{Banfi:2012yh}. Phenomenological predictions for the jet-veto cross section and efficiency are presented in Section~\ref{sec:pheno}. In the final section we summarize our findings.

\section{Factorization and resummation of the cross section}
\label{sec:factorization}

\subsection{Preliminaries}

In the heavy top-quark limit, the effective Lagrangian describing Higgs production via gluon-gluon fusion reads \cite{Inami:1982xt} 
\begin{equation}\label{Leff}
   {\cal L}_{\rm eff} 
   = C_t(m_t^2,\mu)\,\frac{\alpha_s(\mu)}{12\pi}\,\frac{H}{v}\,G_{\mu\nu}^a\,G^{\mu\nu,a} \,,
\end{equation}
where the Wilson coefficient $C_t$ accounts for higher-order loop effects. At NLO in RG-improved perturbation theory, it is given by \cite{Ahrens:2008nc}
\begin{equation}\label{CtNLO}
   C_t(m_t^2,\mu)
   = 1 + \frac{\alpha_s(\mu_t)}{4\pi}\,(5C_A-3C_F) 
    + \frac{\beta_1}{\beta_0}\,\frac{\alpha_s(\mu)-\alpha_s(\mu_t)}{4\pi} + \dots \,,
\end{equation} 
where $\mu_t\sim m_t$ is the matching scale at which the top quark is integrated out, and $\beta_0=11-\frac23\,n_f$ and $\beta_1=102-\frac{38}{3}\,n_f$ (for $N_c=3$ colors) are the first two expansion coefficients of the QCD $\beta$-function. In the presence of a jet veto, the differential cross section for Higgs production at the LHC can then be written as
\begin{equation}\label{sigmaqcd}
\begin{aligned}
   d\sigma(\pTveto) 
   &= \frac{1}{2s} \left( \frac{\alpha_s(\mu)}{12\pi v} \right)^2 C_t^2(m_t^2,\mu)\,
    \frac{d^3q}{(2\pi)^3\,2E_q} \int d^4x\,e^{-iq\cdot x} \\
   &\quad\times{\sum_X}'\,
    \langle P(p_1) P(p_2)|\,G_{\mu\nu}^a\,G^{\mu\nu,a}(x)\,|X\rangle\,
    \langle X|\,G_{\rho\sigma}^b\,G^{\rho\sigma,b}(0)\,|P(p_1) P(p_2)\rangle \,,
\end{aligned}
\end{equation}
where the prime on the sum indicates that we only sum over those hadronic final states $X$ that pass the jet-veto cut. We work with the usual class of sequential recombination jet algorithms \cite{Salam:2009jx}, with distance measure
\begin{equation}\label{algo}
\begin{aligned}
   d_{ij} &= \mbox{min}(p_{Ti}^n,p_{Tj}^n)\,
    \frac{\sqrt{\Delta y_{ij}^2+\Delta\phi_{ij}^2}}{R} \,, \\
   d_{iB} &= p_{Ti}^n \,,
\end{aligned}
\end{equation}
where $n=1$ corresponds to the $k_T$ algorithm \cite{Catani:1993hr,Ellis:1993tq}, $n=0$ to the Cambridge-Aachen algorithm \cite{Dokshitzer:1997in,Wobisch:1998wt}, and $n=-1$ to the anti-$k_T$ algorithm \cite{Cacciari:2008gp}. The particles with the smallest distance are combined into a new ``particle'', whose momentum is the sum of the momenta of the parent particles. If the smallest distance is $d_{iB}$, then particle $i$ is considered a jet and removed from the list. The procedure is iterated until all particles are grouped into jets, i.e., the algorithm is inclusive. The corrections to the cross section which are logarithmically enhanced at small $\pTveto$ arise from emissions that are soft or collinear to the proton beams. In perturbation theory, a sensitivity to the jet algorithm appears first at NNLO through diagrams with two real emissions. At this order a dependence on $R$ arises, but the cross section is still independent of $n$. The two particles are clustered into a single jet whenever $\sqrt{\Delta y_{12}^2+\Delta\phi_{12}^2}<R$, and otherwise they are treated as two jets.

We will analyze the cross section using the formalism of SCET, in which highly energetic particles aligned with the colliding protons are described in terms of collinear and anti-collinear quark and gluon fields, and soft particles emitted from the beam jets are described in terms of soft fields. The effective theory implements an expansion of scattering amplitudes in powers of the small parameter $\lambda\sim\pTveto/m_H$, where the jet veto sets the characteristic size of all transverse momenta in the process. We introduce two light-like reference vectors $n^\mu$ and $\bar n^\mu$ (satisfying $n\cdot\bar n=2$) parallel to the beam axis and decompose all 4-vectors in the light-cone basis spanned by these vectors,
\begin{equation}\label{lightcone}
   p^\mu = n\cdot p\,\frac{\bar n^\mu}{2} + \bar n\cdot p\,\frac{n^\mu}{2} + p_\perp^\mu 
   \equiv p_+^\mu + p_-^\mu + p_\perp^\mu \,.
\end{equation}
The different types of modes relevant to our discussion are characterized by the scalings of their momenta $(p_+,p_-,p_\perp)$ with powers of $\lambda$, namely $p_c^\mu\sim m_H(\lambda^2,1,\lambda)$ for collinear particles, $p_{\bar c}^\mu\sim m_H(1,\lambda^2,\lambda)$ for anti-collinear particles, and $p_s^\mu\sim m_H(\lambda,\lambda,\lambda)$ for soft particles. Hence, the particles in these three categories have transverse momenta of order the jet veto, but very different rapidities.

At leading order in power counting, the gluon operator in (\ref{Leff}) can be matched onto an operator in SCET consisting of gauge-invariant, collinear and anti-collinear gluon fields ${\cal A}_c^\mu$ and ${\cal A}_{\bar c}^\mu$. The formalism of expressing SCET operators in terms of gauge-invariant building blocks is described, e.g., in \cite{Bauer:2002nz,Hill:2002vw}. In the present case, the corresponding matching relation reads \cite{Ahrens:2008nc}
\begin{equation}\label{current}
   G_{\mu\nu}^a\,G^{\mu\nu,a} 
   \to -2q^2\,C_S(-q^2-i\epsilon,\mu)\,g_{\mu\nu}^\perp\,
    {\cal A}_c^{\mu,a} \big( S_n^\dagger S_{\bar n} \big)^{ab} {\cal A}_{\bar c}^{\nu,b} \,,
\end{equation}
where $q^2=m_H^2$ is the time-like momentum transfer carried by the current. The explicit expression for the hard matching coefficient $C_S$ at NLO in RG-improved perturbation theory is given in relation (\ref{CSevolsol}) of the appendix. Note that only the physical (transverse) polarization states of the gluons appear in the above relation. The objects $S_n$ and $S_{\bar n}$ are soft Wilson lines in the directions of the light-like vectors $n$ and $\bar n$, which extend from the interaction point to infinity. Assuming that all quantum fields vanish at infinity, the product $S_n^\dagger S_{\bar n}$ corresponds to a gauge-invariant Wilson loop with a cusp at the point where the two Wilson lines come together. In (\ref{current}) we have already applied the SCET decoupling transformation \cite{Bauer:2001yt}, which removes the interactions between (anti-)collinear and soft fields from the SCET Lagrangian. It is then always possible to group the fields of different types together and factorize any SCET operator into gauge-invariant products of collinear, anti-collinear, and soft fields. Likewise, the final states $|X\rangle$ can be written in the factorized form $|X\rangle=|X_c\rangle\otimes|X_{\bar c}\rangle\otimes|X_s\rangle$. 

\subsection{Jet clustering and mode separation}

We now exploit the fact that the jet clustering algorithm does not group particles with different momentum scalings (collinear, anti-collinear, or soft) into the same jet. The reason is that, generically, the rapidity difference between two such particles is of order $\ln(m_H/\pTveto)$, which is parametrically larger than the radius parameter $R$ employed in the jet algorithm (\ref{algo}). Only in corners of the phase space, e.g.\ when a soft emission becomes collinear to the beam, soft and collinear radiation can be combined. However, since the cross section does not exhibit additional singularities in the corresponding limit, such configurations only give rise to power-suppressed contributions. 

To make this argument more precise, we must first specify the power counting adapted for the radius parameter $R$, which we choose such that
\begin{equation}\label{R1}
   \lambda\ll R\ll\ln\frac{1}{\lambda} \,.
\end{equation}
If $R$ was too large, the argument about clustering given above (and explained in more detail below) would fail, while for too small $R$ large logarithms $\ln^n\!R$ would arise, which would require a special treatment. These ``clustering logarithms'' have a complicated structure in higher orders \cite{Kelley:2012kj,Kelley:2012zs}, and it is currently not understood how to resum them. Assuming that $R$ satisfies relation (\ref{R1}), we will derive a factorization theorem valid at leading power in $\lambda$ in the formal limit where $\lambda\ll 1$. Once the result has been established, it will hold for any value $R={\cal O}(1)$. Even though $\lambda$ is not very small in practice, the above condition is satisfied for realistic values $R=0.4$ or 0.5 used in experimental analyses. For instance, with $\lambda=\pTveto/m_H$ we have $\lambda\approx 0.16$ and $\ln(1/\lambda)\approx 1.83$ for $\pTveto=20$\,GeV and $m_H=125$\,GeV. Our resummed results for the cross section will receive power corrections in $\lambda$, which will be added when we match them to fixed-order perturbation theory. 

Consider now the behavior of soft, collinear, and anti-collinear modes under the jet clustering algorithm (\ref{algo}). For two modes from the same sector, the quantities $d_{ij}$ and $d_{iB}$ are parametrically of the same order, and depending on the precise values of the transverse momenta, rapidities, and polar angles of the two particles, they are either clustered together or considered two jets. On the other hand, for two modes from different sectors (e.g.\ a soft and a collinear particle) we have parametrically $d_{ij}\sim\lambda^n\ln(1/\lambda)$ but $d_{iB}\sim\lambda^n$, where it is crucial that $R={\cal O}(1)$ or smaller. It follows that {\em generically\/} the two particles are not clustered into the same jet. Since the soft and (anti-)collinear modes have the same virtuality, they live along a hyperbola in the $(p_+,p_-)$ plane, and their precise separation along this hyperbola is to some extent arbitrary. The fact that these modes differ by large rapidities gives rise to large logarithms, which are accounted for by the collinear anomaly. In complete analogy with the construction of the SCET Lagrangian, where based on the {\em generic\/} scalings of the fields one does not include soft-collinear interaction terms, it is unnecessary to consider the degenerate case where a collinear and a soft mode near the boundary are clustered into a single jet. Since there are no enhancements of the cross section in these power-suppressed phase-space regions, boundary effects do not contribute at leading power.

One may also worry about the contributions from modes with smaller virtualities, $p^2\ll(\pTveto)^2$. For example, an on-shell soft mode, which accidentally is closely aligned with the beam axis, would have momentum scaling $(\lambda^2,\lambda,\lambda^{3/2})$. This mode has the same rapidity as a collinear mode, and hence it may be clustered with such a mode. Indeed, it may also be regarded as a collinear mode, whose minus component is accidentally small. The important point is that, because of their small transverse momenta, such modes play no role for the total $p_T$ of a jet. Therefore, an arbitrary number of them can be emitted, and their effect cancels out in the factorization theorem. This is in analogy with the cancellation of ultra-soft modes in the factorization theorem for the Drell-Yan cross section at small transverse momentum~\cite{Becher:2010tm}.

It follows from this discussion that, at leading power in $\lambda$, the cross section factorizes into the production of soft and (anti-)collinear jets, and the jet veto can be applied separately in each sector. Changing variables from $d^3q$ to $dy\,d^2q_\perp$, where $q_\perp$ and $y$ denote the transverse momentum and rapidity of the Higgs boson, we obtain from (\ref{sigmaqcd})
\begin{equation}
\begin{aligned}
   d\sigma(\pTveto) 
   &= \frac{1}{2s} \left( \frac{\alpha_s(\mu)}{12\pi v} \right)^2 C_t^2(m_t^2,\mu)\,
    4 m_H^4 \left| C_S(-m_H^2,\mu) \right|^2\,\frac{dy}{4\pi}\,\frac{d^2q_\perp}{(2\pi)^2} 
    \int d^4x\,e^{-iq\cdot x} \\
   &\quad\times{\sum_{X_c}}'\,\langle P(p_1)|\,{\cal A}_c^{\mu,a}(x_+ +x_\perp)\,|X_c\rangle\,
    \langle X_c|\,{\cal A}_c^{\nu,d}(0)\,|P(p_1)\rangle\,g_{\mu\rho}^\perp\,g_{\nu\sigma}^\perp 
    \\[-2mm]
   &\quad\times{\sum_{X_{\bar c}}}'\,
    \langle P(p_2)|\,{\cal A}_{\bar c}^{\rho,b}(x_- +x_\perp)\,|X_{\bar c}\rangle\,
    \langle X_{\bar c}|\,{\cal A}_{\bar c}^{\sigma,c}(0)\,|P(p_2)\rangle \\[-2mm]
   &\quad\times{\sum_{X_s}}'\,
    \langle\,0\,|\,\big( S_n^\dagger S_{\bar n} \big)^{ab}(x_\perp)\,|X_s\rangle\,
    \langle X_s|\,\big( S_{\bar n}^\dagger S_n \big)^{cd}(0)\,|\,0\,\rangle + \dots \,,
\end{aligned}
\end{equation}
where the dots represent power-suppressed terms in the SCET expansion parameter $\lambda\sim\pTveto/m_H$, which we neglect. Above, we have already performed the multipole expansion of the SCET fields, which neglects those components of $x$ in the arguments of the fields that would introduce subleading momentum components \cite{Beneke:2002ph}. To identify these components we have used that $x^\mu\sim(1,1,\lambda^{-1})$, which is conjugate to the momentum $q^\mu\sim m_H(1,1,\lambda)$ of the Higgs boson.

We now use color conservation for the individual matrix elements and define
\begin{eqnarray}\label{BSdef}
   {\cal B}_c^{\mu\nu}(z,x_\perp,\pTveto,\mu)
   &=& - \frac{z\,\bar n\cdot p}{2\pi} \int dt\,e^{-izt\bar n\cdot p}\,
    {\sum_{X_c}}'\,\langle P(p)|\,{\cal A}_{c\perp}^{\mu,a}(t\bar n+x_\perp)\,|X_c\rangle\,
    \langle X_c|\,{\cal A}_{c\perp}^{\nu,a}(0)\,|P(p)\rangle \,, \nonumber\\[-1mm]
   {\cal S}(x_\perp,\pTveto,\mu)
   &=& \frac{1}{N_c^2-1}\,{\sum_{X_s}}'\,
    \langle\,0\,|\,\big( S_n^\dagger S_{\bar n} \big)^{ab}(x_\perp)\,|X_s\rangle\,
    \langle X_s|\,\big( S_{\bar n}^\dagger S_n \big)^{ba}(0)\,|0\rangle \,.
\end{eqnarray}
The function ${\cal B}_c^{\mu\nu}$ describes the structure of the jet of collinear particles inside one of the colliding protons (the one moving along the light-like direction $n^\mu$), which is probed at small transverse distances $x_\perp$. In the context of SCET, such functions are referred to as beam functions \cite{Stewart:2009yx}. The corresponding function ${\cal B}_{\bar c}^{\mu\nu}$ for the other beam jet is given by the same formula with the replacements $\bar n\to n$ and $c\to\bar c$. The soft function ${\cal S}$ describes the physics of soft gluons emitted from the colliding beam particles. Note that both the beam functions and the soft function depend on the jet algorithm used for the jet veto. This is implicit in our notation, which only makes the dependence on $\pTveto$ explicit.

When inverting the first relation in (\ref{BSdef}), one must take into account that the beam function ${\cal B}_c^{\mu\nu}(z,x_\perp,\pTveto,\mu)$ has support for $-1\le z\le 1$, and it can be shown that ${\cal B}_c^{\mu\nu}(-z,x_\perp,\pTveto,\mu)=-{\cal B}_c^{\mu\nu}(z,-x_\perp,\pTveto,\mu)$ (see e.g.\ \cite{Collins:2011zzd}). Restricting the integration range to positive $z$ values thus leads to an extra factor of 2 for each beam function. After some straightforward algebra, we obtain 
\begin{equation}\label{sig1}
\begin{aligned}
   d\sigma(\pTveto) 
   &= \sigma_0(\mu)\,C_t^2(m_t^2,\mu) \left| C_S(-m_H^2,\mu) \right|^2\,\frac{m_H^2}{\tau s}\,
    dy\,\frac{d^2q_\perp}{(2\pi)^2} \int d^2x_\perp\,e^{-iq_\perp\cdot x_\perp} \\
   &\quad\times 2 {\cal B}_c^{\mu\nu}(\xi_1,x_\perp,\pTveto,\mu)\,
    {\cal B}_{\bar c\,\mu\nu}(\xi_2,x_\perp,\pTveto,\mu)\,
    {\cal S}(x_\perp,\pTveto,\mu) \,,
\end{aligned}
\end{equation}
where $\xi_{1,2}=\sqrt{\tau}\,e^{\pm y}$ and $\tau=(m_H^2+|q_\perp^2|)/s$. The Born-level cross section is
\begin{equation}
   \sigma_0(\mu) = \frac{m_H^2\,\alpha_s^2(\mu)}{72\pi (N_c^2-1) s v^2} \,.
\end{equation}

Up to corrections of order $\lambda^2$, one can replace $\tau\to m_H^2/s$ in (\ref{sig1}). Given that the peak of the transverse-momentum distribution is around 10\,GeV \cite{Bozzi:2007pn}, this is a very good approximation. Integrating over $q_\perp$, we then obtain the simple result
\begin{equation}
\begin{aligned}
   \frac{d\sigma(\pTveto)}{dy}
   &= \sigma_0(\mu)\,C_t^2(m_t^2,\mu) \left| C_S(-m_H^2,\mu) \right|^2 \\
   &\quad\times 2{\cal B}_c^{\mu\nu}(\xi_1,0,\pTveto,\mu)\,
    {\cal B}_{\bar c\,\mu\nu}(\xi_2,0,\pTveto,\mu)\,{\cal S}(0,\pTveto,\mu) \,.
\end{aligned}
\end{equation}
Since the jet veto does not prefer any direction in the transverse plane, we can replace
\begin{equation}
   {\cal B}_c^{\mu\nu}(z,0,\pTveto,\mu) = \frac{g_\perp^{\mu\nu}}{2}\,{\cal B}_c(z,\pTveto,\mu) \,,
\end{equation}
where
\begin{equation}
   {\cal B}_c(z,\pTveto,\mu)
   = - \frac{z\,\bar n\cdot p}{2\pi} \int dt\,e^{-izt\bar n\cdot p}\,
    {\sum_{X_c}}'\,\langle P(p)|\,{\cal A}_{c\perp}^{\mu,a}(t\bar n)\,|X_c\rangle\,
    \langle X_c|\,{\cal A}_{c\perp\mu}^a(0)\,|P(p)\rangle \,.
\end{equation}
Note the similarity of this function with the usual definition of the gluon distribution function in the proton, which in SCET notation reads \cite{Becher:2010tm}
\begin{equation}
   \phi_{g/P}(z,\mu)
   = - \frac{z\,\bar n\cdot p}{2\pi} \int dt\,e^{-izt\bar n\cdot p}\,
    \langle P(p)|\,{\cal A}_{c\perp}^{\mu,a}(t\bar n)\,{\cal A}_{c\perp\mu}^a(0)\,|P(p)\rangle \,.
\end{equation}
Dropping the vanishing argument $x_\perp=0$ in the soft function, we finally obtain
\begin{equation}\label{dsigdy}
   \frac{d\sigma(\pTveto)}{dy}
   = \sigma_0(\mu)\,C_t^2(m_t^2,\mu) \left| C_S(-m_H^2,\mu) \right|^2 
    {\cal B}_c(\xi_1,\pTveto,\mu)\,{\cal B}_{\bar{c}}(\xi_2,\pTveto,\mu)\,
    {\cal S}(\pTveto,\mu) \,.
\end{equation}

\subsection{Collinear factorization anomaly}

The above formula appears to disentangle the hard scales $m_t$ and $m_H$ from the soft scale $\pTveto\ll m_H$. The factorization is however not yet complete. Closer inspection shows that the individual ingredients to the cross section (\ref{dsigdy}) are ill-defined in dimensional regularization, so that an additional regularization is needed in an intermediate step. The situation closely resembles the case of the Drell-Yan cross section at small transverse momentum analyzed in \cite{Becher:2010tm}. As in this case, it is convenient to employ an analytic regulator for this purpose. 

It was recently shown that the additional analytic regularization is only needed for the phase space, and can be achieved by replacing the usual phase-space integration by \cite{Becher:2011dz} 
\begin{equation}\label{anal}
   \int\!d^dk\,\delta(k^2)\,\theta(k^0) \;\;\to\;\; 
   \int\!d^dk \left( \frac{\nu}{n\cdot k} \right)^\alpha \delta(k^2)\,\theta(k^0) \,.
\end{equation}
The regularization softens light-cone singularities in the cross section. It introduces a new scale $\nu$, which plays an analogous role to the scale $\mu$ entering in dimensional regularization. In the above form, the analytic regularization leaves the structure of the effective theory intact, and all steps in the derivation of the factorization formula (\ref{dsigdy}) remain the same, except that the phase-space integrals implicit in the sums over intermediate states $\sum_X'$ in (\ref{BSdef}) must be regularized according to (\ref{anal}). Note that the symmetry $n\leftrightarrow\bar n$ between the collinear and anti-collinear beam-jet functions is broken in the presence of the regulator, which involves the $n\cdot k$ component in both sectors. An alternative analytic regularization scheme, based on a regularization of the SCET Wilson lines, was proposed in \cite{Chiu:2011qc,Chiu:2012ir}.

In the presence of the extra regulator, the jet and soft functions are separately well defined, but individually they contain singularities ($1/\alpha^n$ poles) in the analytic regulator. These cancel in the product of the jet and soft functions entering a physical cross section, so that the regulator can be removed at the end of the analysis. Importantly, however, the $\alpha$ regulator breaks the invariance of the soft and jet functions under a rescaling of the large light-cone components of the momenta of the external particles. This symmetry of the low-energy theory is anomalous: it is not recovered after the regulator is removed. Instead, one finds an anomalous dependence of the low-energy theory on the hard scale $m_H$. Following \cite{Becher:2010tm,Becher:2011pf}, we now derive the all-order structure of this dependence by solving differential equations expressing the independence of the cross section of the analytic regularization scale $\nu$. 

In order to constrain the anomaly, we first slightly generalize our observable. We veto an event if it contains left-moving jets with transverse momentum above a value $\pLveto$ or right-moving jets with transverse momentum above $\pRveto$. The beam-jet functions are the same as before, with $\pTveto$ replaced by $\pLveto$ ($\pRveto$) for the left-moving (right-moving) jet. The soft function, on the other hand, now depends both on $\pLveto$ and $\pRveto$, since soft jets can be emitted in any direction. Other than this, the form of the cross section is exactly the same as above. Our previous result (\ref{dsigdy}) is recovered in the special case where $\pLveto=\pRveto=\pTveto$. Choosing the values differently will, however, allow us to derive additional constraints on the form of the anomaly. 

Since the original QCD expression (\ref{sigmaqcd}) does not require analytic regularization, we are guaranteed that the limit $\alpha\to 0$ can be taken in the product of the jet and soft functions in (\ref{dsigdy}). However, upon expanding in the regulator, the individual functions involve logarithms of the scale $\nu$. Taking the logarithm of the product $P={\cal B}_c\,{\cal B}_{\bar c}\,{\cal S}$ of jet and soft functions in (\ref{dsigdy}), we have
\begin{equation}\label{lnP}
   \ln P = \ln{\cal B}_c\Big(\!\ln\frac{m_H\nu}{\mu^2};\xi_1,\pLveto,\mu\Big)
    + \ln{\cal B}_{\bar c}\Big(\!\ln\frac{\nu}{m_H};\xi_2,\pRveto,\mu\Big) 
    + \ln{\cal S}\Big(\!\ln\frac{\nu}{\mu};\pLveto,\pRveto,\mu\Big) \,.
\end{equation}
For $\mu\sim\pLveto\sim\pRveto$, no large logarithms beyond the ones we display explicitly can arise. Their functional form reflects the scaling of $n\cdot k$ in the various momentum regions, which was given after (\ref{lightcone}). The structure of this result closely resembles that encountered in our analysis of jet broadening presented in \cite{Becher:2011pf}. The fact that $\ln P$ as well as its derivatives with respect to $m_H$ must be independent of $\nu$ gives rise to a set of simple differential equations, which imply that the three component functions can be at most quadratic in their first argument, such that 
\begin{equation}
\begin{aligned}
   \ln{\cal B}_c\Big(\!\ln\frac{m_H\nu}{\mu^2};\xi_1,\pLveto,\mu\Big) 
   &= f_2^{(L)}(\xi_1,\pLveto,\mu)\,\ln^2\frac{m_H\nu}{\mu^2} \\
   &\quad\mbox{}+ f_1^{(L)}(\xi_1,\pLveto,\mu)\,\ln\frac{m_H\nu}{\mu^2}
    + f_0^{(L)}(\xi_1,\pLveto,\mu) \,,
\end{aligned}
\end{equation}
and analogously for the right beam-jet function and the soft function, with expansion coefficients $f_i^{(R)}$ and $f_i^{(S)}$ and logarithms $\ln(\nu/m_H)$ and $\ln(\nu/\mu)$, respectively. The coefficients of the logarithmic terms are furthermore subject to the constraints
\begin{equation}
\begin{aligned}
   &f_1^{(L)}(\xi_1,\pLveto,\mu) + f_1^{(R)}(\xi_2,\pRveto,\mu) 
    + f_1^{(S)}(\pLveto,\pRveto,\mu) = 0 \,, \\
   &f_2^{(L)}(\xi_1,\pLveto,\mu) = f_2^{(R)}(\xi_2,\pRveto,\mu) 
    = - \frac12\,f_2^{(S)}(\pLveto,\pRveto,\mu) \,.
\end{aligned}
\end{equation}
Since the arguments in the various functions are different, it follows that $f_1^{(L,R)}$ and $f_2^{(L,R)}$ must be independent of $\xi_1$ and $\xi_2$. Moreover, the second constraint implies that the coefficients of the quadratic terms can only be a function of $\mu$,
\begin{equation}
   f_2^{(L)}(\xi_1,\pLveto,\mu) \equiv f_2(\mu) \,.
\end{equation}
For the first-order coefficients, the symmetry of $\ln P$ under $L\leftrightarrow R$ and $\xi_1\leftrightarrow\xi_2$ implies that
\begin{equation}
   f_1^{(R)}(\xi_2,\pRveto,\mu) - f_1^{(L)}(\xi_1,\pLveto,\mu) 
   \equiv f_1(\pLveto,\mu) + f_1(\pRveto,\mu) \,,
\end{equation}
and
\begin{equation}
\begin{aligned}
   &f_0^{(L)}(\xi_1,\pLveto,\mu) + f_0^{(R)}(\xi_2,\pRveto,\mu)
    + f_0^{(S)}(\pLveto,\pRveto,\mu) \\
   &\equiv f_0(\xi_1,\pLveto,\mu) + f_0(\xi_2,\pRveto,\mu)
    + \bar f_0^{(S)}(\pLveto,\pRveto,\mu) \,.
\end{aligned}
\end{equation}
This last relation expresses that the dependence on $\xi_1$ and $\xi_2$ is governed by a single function $f_0$, and a possible difference in the functional forms of $f_0^{(L)}$ and $f_0^{(R)}$ can be absorbed into the soft function $f_0^{(S)}$. Using these results, we obtain
\begin{equation}
\begin{aligned}
   \ln P &= 2 f_2(\mu)\,\ln^2\frac{m_H}{\mu}
     - \ln\frac{m_H}{\mu} \left[ f_1(\pLveto,\mu) + f_1(\pRveto,\mu) \right] \\
    &\quad\mbox{}+ f_0(\xi_1,\pLveto,\mu) + f_0(\xi_2,\pRveto,\mu) 
     + \bar f_0^{(S)}(\pLveto,\pRveto,\mu) \,.
\end{aligned}
\end{equation}

Since the cross section is scale independent, the derivative of the above expression with respect to the renormalization scale must be equal and opposite to the anomalous dimension of the hard function, which is given in the appendix. This implies
\begin{equation}\label{Prge}
   \frac{d}{d\ln\mu}\,\ln P
   =  - 4\Gamma_{\rm cusp}^A(\mu)\,\ln\frac{m_H}{\mu} - 4\gamma^g(\mu) \,,
\end{equation}
where $\Gamma_{\rm cusp}^A$ is the cusp anomalous dimension in the adjoint representation, and $2\gamma^g=\gamma^S+\gamma^t+\beta(\alpha_s)/\alpha_s$ is the anomalous dimension of the gluon field as defined in \cite{Becher:2009qa}. Since this right-hand side of (\ref{Prge}) is linear in $\ln(m_H/\mu)$, we must require that
\begin{equation}
   \frac{d}{d\ln\mu}\,f_2(\mu) = 0 \,.
\end{equation}
But $f_2(\mu)$ vanishes at tree level, and this equation then implies that it is zero to all orders in perturbation theory, as there is no way to compensate the $\mu$ dependence of the coupling constant. 

The outcome of the above considerations is that the logarithm of the product of beam-jet and soft functions is linear in $\ln(m_H/\mu)$. This result is important for the resummation of logarithms. It implies that in the case $\pLveto=\pRveto=\pTveto$ the product can be written in the form
\begin{equation}\label{refact}
\begin{aligned}
   &\left[ {\cal B}_c\Big(\!\ln\frac{m_H\nu}{\mu^2};\xi_1,\pLveto,\mu\Big)\,
    {\cal B}_{\bar c}\Big(\!\ln\frac{\nu}{m_H};\xi_2,\pRveto,\mu\Big)\, 
    {\cal S}\Big(\!\ln\frac{\nu}{\mu};\pLveto,\pRveto,\mu\Big) \right]_{q^2=m_H^2} \\
   &= \left( \frac{m_H}{\pLveto} \right)^{-F_{gg}(\pLveto,\mu)}
    \left( \frac{m_H}{\pRveto} \right)^{-F_{gg}(\pRveto,\mu)}
    B_g(\xi_1,\pLveto,\mu)\,B_g(\xi_2,\pRveto,\mu)\,S_g(\pLveto,\pRveto,\mu) \,,
\end{aligned}
\end{equation}
where we have defined
\begin{equation}
\begin{aligned}
   F_{gg}(\pTveto,\mu) &= f_1(\pTveto,\mu) \,, \\[1mm]
   \ln B_g(\xi,\pTveto,\mu) 
   &= f_1(\pTveto,\mu)\,\ln\frac{\mu}{\pTveto} + f_0(\xi,\pTveto,\mu) \,, \\[-1mm]
   \ln S_g(\pLveto,\pRveto,\mu) &= \bar f_0^{(S)}(\pLveto,\pRveto,\mu) \,.
\end{aligned}
\end{equation}
The $m_H$ dependence due to the collinear anomaly is now explicit in (\ref{refact}). Note also that the dependence on the longitudinal variables $\xi_1$ and $\xi_2$ is factorized and given in terms of a single function $B_g(\xi,\pTveto,\mu)$. The definition of these functions is not unique, since it is possible to shift $\pTveto$-dependent terms into the soft function. For the special case $\pLveto=\pRveto=\pTveto$, it is therefore natural to absorb the square root of the soft function into the definition of the beam-jet function $B_g$.

The structure of the result for the cross section, obtained when inserting (\ref{refact}) into (\ref{dsigdy}), is particularly simple for the case of the jet veto. In previous applications of the collinear anomaly in the context of small-$q_T$ resummation for Drell-Yan production \cite{Becher:2010tm} and jet broadening in $e^+ e^-$ annihilations \cite{Becher:2011pf}, the anomalous exponents analogous to $F_{gg}$ depended on some convolution variables, which were shared with the jet and soft functions in the factorization theorem. After integration over these variables, the resulting dependence on the hard scale can be quite complicated (see in particular the discussion in \cite{Becher:2011xn}). In the present case, on the other hand, the anomaly leads to an extra term in the cross section which, at fixed $\pTveto$, is a pure power of $m_H$. 

\subsection{Short-distance expansion of the beam-jet function}

For the final factorization step, we now use that in practice the jet veto is much larger than the scale $\Lambda_{\rm QCD}$ governing non-perturbative hadronic effects in QCD. It is thus possible to calculate the physics associated with $\pTveto$ in perturbation theory, and to relate the refactorized beam-jet function $B_g\,S_g^{1/2}$ to conventional PDFs \cite{Stewart:2009yx,Collins:1984kg}. We write this relation in the form
\begin{equation}\label{pdfmatch}
   B_g(\xi,\pTveto,\mu)\,S_g^{1/2}(\pTveto,\pTveto,\mu)
   = \sum_{i=g,q,\bar q} \int_\xi^1\!\frac{dz}{z}\,
    I_{g\leftarrow i}(z,\pTveto,\mu)\,\phi_{i/P}(\xi/z,\mu) \,,
\end{equation}
which is valid up to hadronic corrections suppressed by powers of $\Lambda_{\rm QCD}/\pTveto$. By means of (\ref{refact}) and (\ref{pdfmatch}), the cross section in (\ref{dsigdy}) can now be expressed in the final form
\begin{eqnarray}\label{dsigfinal}
   \frac{d\sigma(\pTveto)}{dy}
   &=& \sigma_0(\mu)\,C_t^2(m_t^2,\mu) \left| C_S(-m_H^2,\mu) \right|^2 
    \left( \frac{m_H}{\pTveto} \right)^{-2F_{gg}(\pTveto,\mu)} \\[-1mm]
   &&\times \sum_{i,j} 
    \int_{\xi_1}^1\!\frac{dz_1}{z_1} \int_{\xi_2}^1\!\frac{dz_2}{z_2}\,
    I_{g\leftarrow i}(z_1,\pTveto,\mu)\,I_{g\leftarrow j}(z_2,\pTveto,\mu)\,
    \phi_{i/P}(\xi_1/z_1,\mu)\,\phi_{j/P}(\xi_2/z_2,\mu) \,, \nonumber
\end{eqnarray}
where the sums over $i,j$ extend over all flavors of partons (gluons, quarks, and anti-quarks). In the above factorization theorem, the dependence on the two disparate scales $m_H$ and $\pTveto\ll m_H$ is completely explicit. Large logarithms of their ratio can be resummed by choosing the factorization scale around the value of the jet veto, $\mu\sim\pTveto$. With such a choice, the functions $F_{gg}$ and $I_{g\leftarrow i}$ have well-behaved perturbative expansions free of large logarithms. Explicit expressions for these functions will be presented in the next section. Large logarithms of $m_H/\pTveto$ are contained in the anomaly term (in exponentiated form) and in the short-distance matching coefficients $C_t$ and $C_S$, whose expressions are known at NNLO in RG-improved perturbation theory. They are collected in the appendix. 

A particularly nice formula is obtained after we integrate the differential cross section over rapidity. It reads
\begin{equation}\label{sigfinal}
\begin{aligned}
   \sigma(\pTveto) 
   &= \sigma_0(\mu)\,C_t^2(m_t^2,\mu) \left| C_S(-m_H^2,\mu) \right|^2 
    \left( \frac{m_H}{\pTveto} \right)^{-2F_{gg}(\pTveto,\mu)} \\[-1mm]
   &\quad\times\sum_{i,j=g,q} \int_\tau^1\!\frac{dz}{z}\,
    \II_{ij}(z,\pTveto,\mu)\,\ff_{ij}(\tau/z,\mu) \,,
\end{aligned}
\end{equation}
where the Mellin convolution $\II_{ij}\equiv I_{g\leftarrow i}\otimes I_{g\leftarrow j}$ of two kernel functions is defined as
\begin{equation}
   \II_{ij}(z,\pTveto,\mu) 
   = \int_z^1\!\frac{du}{u}\,
    I_{g\leftarrow i}(u,\pTveto,\mu)\,I_{g\leftarrow j}(z/u,\pTveto,\mu) \,; \quad i,j=g,q \,,
\end{equation}
while $\ff_{ij}$ are flavor-singlet parton luminosity functions, given by $\ff_{gg}=\phi_{g/P}\otimes\phi_{g/P}$, $\ff_{gq}=\phi_{g/P}\otimes\sum_q\left(\phi_{q/P}+\phi_{\bar q/P}\right)$, etc.

To complete our discussion, we present the RG equations obeyed by the various component functions in (\ref{dsigfinal}). The evolution equation for the Wilson coefficient $C_t$ was first given in \cite{Inami:1982xt}, while that for $C_S$ can be found, e.g., in \cite{Ahrens:2008nc}. For completeness, we list the corresponding RG equations in the  appendix. The evolution equations for the functions $F_{gg}$ and $I_{g\leftarrow i}$ then follow from the scale invariance of the cross section. In analogy with our discussion in \cite{Becher:2010tm}, we obtain
\begin{equation}\label{evol}
\begin{aligned}
   \frac{d}{d\ln\mu}\,F_{gg}(\pTveto,\mu) 
   &= 2\Gamma_{\rm cusp}^A(\mu) \,, \\[-1mm]
   \frac{d}{d\ln\mu}\,I_{g\leftarrow i}(z,\pTveto,\mu)
   &= \Big[ 2\Gamma_{\rm cusp}^A(\mu)\,\ln\frac{\mu}{\pTveto} - 2\gamma^g(\mu) \Big]\,
    I_{g\leftarrow i}(z,\pTveto,\mu) \\
   &\quad\mbox{}- \sum_j \int_z^1\!\frac{du}{u}\,I_{g\leftarrow j}(u,\pTveto,\mu)\, 
    {\cal P}_{j\leftarrow i}(z/u,\mu) \,,
\end{aligned}
\end{equation}
where ${\cal P}_{j\leftarrow i}$ are the usual DGLAP splitting functions.

\section{Resummation at NNLL order}
\label{sec:oneloop}

For a consistent treatment at NLO in RG-improved perturbation theory, we require the anomaly exponent $F_{gg}$ at two-loop order, while the one-loop expressions for the kernel functions $I_{g\leftarrow i}$ suffice. Up to ${\cal O}(\alpha_s^2)$, the general solution to the first evolution equation in (\ref{evol}) reads \cite{Becher:2010tm}
\begin{equation}\label{Fgg2}
   F_{gg}(\pTveto,\mu)
   = a_s \left( \Gamma_0^A\,L_\perp + d_1^{\rm veto}\hspace{-0.5mm} \right)
    + a_s^2 \left( \Gamma_0^A\beta_0\,\frac{L_\perp^2}{2}
    + \Gamma_1^A\,L_\perp + \dv \right) ; \quad
    L_\perp = 2\ln\frac{\mu}{\pTveto} \,,
\end{equation}
where $a_s\equiv\alpha_s(\mu)/(4\pi)$, $\Gamma_0^A=4C_A$ and $\Gamma_1^A=\big(\frac{268}{9}-\frac{4\pi^2}{3}\big)\,C_A^2-\frac{80}{9}\,C_A T_F n_f$ are the one- and two-loop coefficients of the cusp anomalous dimension, and $\beta_0=\frac{11}{3}\,C_A-\frac43\,T_F n_f$. The constant terms $d_1^{\rm veto}\hspace{-0.5mm}$ and $\dv$ are independent of the jet veto $\pTveto$ but in general can depend on the jet algorithm. At one-loop order, the solution to the second evolution equation in (\ref{evol}) can be written as \cite{Becher:2010tm}
\begin{equation}\label{Ires}
   I_{g\leftarrow i}(z,\pTveto,\mu) 
   = \delta(1-z)\,\delta_{gi} \left[ 1 + a_s \left( \Gamma_0^A\,\frac{L_\perp^2}{4}
    - \gamma_0^g\,L_\perp \right) \right]
    + a_s \left[ - {\cal P}_{g\leftarrow i}^{(1)}(z)\,\frac{L_\perp}{2}
    + {\cal R}_{g\leftarrow i}(z) \right] , 
\end{equation}
where $\gamma_0^g=-\beta_0$, 
\begin{equation}
\begin{aligned}
   {\cal P}_{g\leftarrow g}^{(1)}(z) 
   &= 8C_A \left[ \frac{z}{\left(1-z\right)_+} + \frac{1-z}{z} + z(1-z) \right]
    + 2\beta_0\,\delta(1-z) \,, \\
   {\cal P}_{g\leftarrow q}^{(1)}(z) 
   &= 4C_F\,\frac{1+(1-z)^2}{z}
\end{aligned}
\end{equation}
are the one-loop DGLAP splitting functions, and the remainder functions ${\cal R}_{g\leftarrow i}(z)$ collect terms that are independent of $\pTveto$. 

Up to the order shown in the equations above, the functions $F_{gg}$ and $I_{g\leftarrow i}$ entering the factorization formulae (\ref{dsigfinal}) and (\ref{sigfinal}) are closely related to corresponding quantities entering the resummed cross section for Higgs production at small transverse momentum \cite{Becher:2010tm,inprep}. The only difference is that in that case $L_\perp=\ln(x_T^2\mu^2/b_0^2)$ with $b_0=2 e^{-\gamma_E}$. An exception is the two-loop coefficient $\dv$, which receives contributions from diagrams with two real emissions and is thus sensitive to the definition of the jet veto.

\begin{figure}
\begin{center}
\begin{tabular}{cccc}
\includegraphics[height=0.132\textwidth]{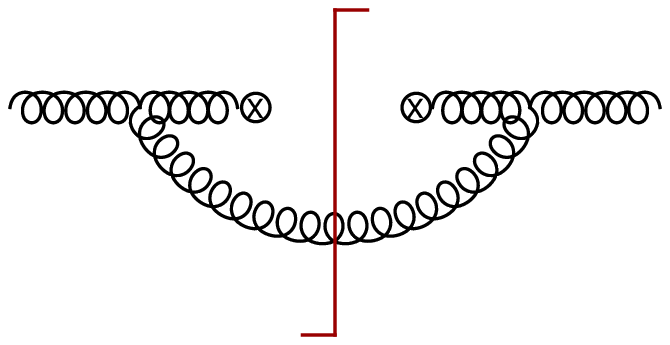} & 
\includegraphics[height=0.132\textwidth]{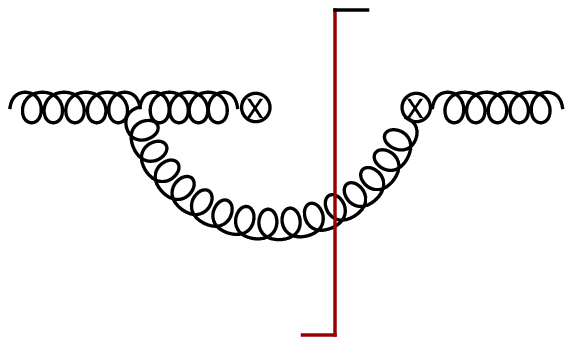} & 
\includegraphics[height=0.132\textwidth]{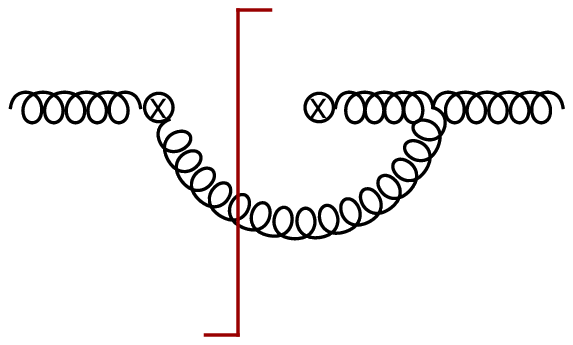} & 
\includegraphics[height=0.132\textwidth]{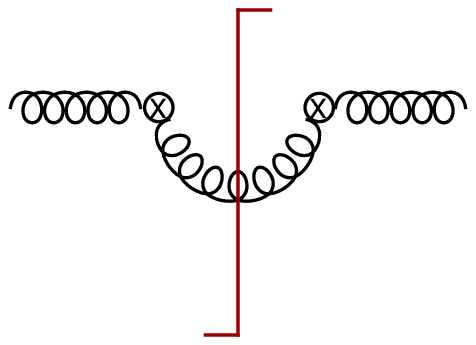} \\[0.1cm]
\end{tabular}
\includegraphics[height=0.132\textwidth]{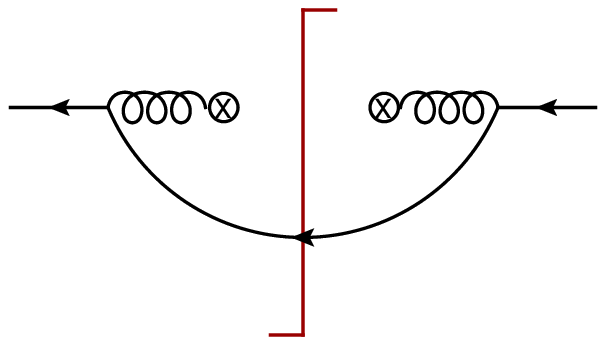}
\vspace{-0.3cm}
\caption{\label{fig:graphs}
One-loop diagrams contributing to the anomaly exponent $F_{gg}$ and the matching coefficients $I_{g\leftarrow g}$ (top row) and $I_{g\leftarrow q}$ (bottom row). Vertical lines indicate cut propagators.}
\end{center}
\end{figure}

The calculation of the kernel functions at one-loop order proceeds in close analogy to the calculation of the corresponding functions entering in the small-$q_T$ resummation for Drell-Yan or Higgs production, which was performed in SCET in \cite{Becher:2010tm,inprep}. The notation used above matches that adopted in these papers. The relevant Feynman graphs are shown in Figure~\ref{fig:graphs}. Since at this order only a single parton line is cut (corresponding to one-particle states $X$), the jet veto is trivially implemented as an upper cut on the transverse momentum of that parton. In addition to the collinear one-loop corrections shown in Figure~\ref{fig:graphs}, the corrections to the soft function in (\ref{BSdef}) must be computed. At one-loop order this is trivial, since the relevant integrals are scaleless and therefore vanish in our regularization scheme. The results of our calculation agree with the above expressions, which were derived from the corresponding evolution equations. For the one-loop constant term in (\ref{Fgg2}) and the remainder functions in (\ref{Ires}), we obtain
\begin{equation}
   d_1^{\rm veto}\hspace{-0.5mm} = 0 \,, \qquad
   {\cal R}_{g\leftarrow g}(z) = - C_A\,\frac{\pi^2}{6}\,\delta(1-z) \,, \qquad
   {\cal R}_{g\leftarrow q}(z) = 2C_F z \,.
\end{equation}
The soft function ${\cal S}$ vanishes at one-loop order (like in the case of small-$q_T$ resummation for Drell-Yan production \cite{Becher:2010tm}), but it is expected to be non-zero in higher orders, where the jet veto imposes non-trivial phase-space constraints. 

The two-loop coefficient $\dv$ in (\ref{Fgg2}) cannot be determined from our one-loop analysis. It accounts for genuine two-loop effects and is therefore different from the corresponding coefficient $d_2^g$ relevant for small-$q_T$ resummation in Higgs production \cite{Becher:2010tm,inprep}. The authors of \cite{Banfi:2012yh} have derived a partial expression for the $R$-dependent piece of the jet-veto cross section at two-loop order in perturbation theory, finding that
\begin{equation}\label{fromGiulia}
   \frac{\sigma(\pTveto)}{\sigma_0(\mu)} \Big|_{\rm 2-loop}
   = 64 C_A\,a_s^2 \left[ f(R)\,\ln\frac{m_H}{\pTveto} + \dots \right] ,
\end{equation} 
where the dots represent $R$-independent terms  and other, possibly $R$-dependent terms that are not enhanced by the large logarithm $\ln(m_H/\pTveto)$. They observed that the function $f(R)$ has a smooth behavior in the limit where $R\to\infty$ and adopted a normalization in which it vanishes like $1/R$ in this limit \cite{privcom}. For very large $R$, the jet clustering algorithm combines all particles into a single jet, since $d_{ij}\to 0$ in (\ref{algo}). The jet veto then implies an upper bound on the vector sum of the transverse momenta of all particles, which is equivalent to a bound on the transverse momentum of the Higgs boson. Assuming that the formalism of \cite{Banfi:2012yh} is still valid\footnote{See the note added in proof.} 
in the limit of very large $R$, we can then match the $R$-dependent two-loop terms obtained in this paper with the ones following from (\ref{sigfinal}) and from the results obtained in our previous work \cite{Becher:2010tm,inprep} for the transverse-momentum spectrum of the Higgs boson. In this way, we obtain
\begin{equation}\label{d2v}
   \dv = d_2^g + 2\zeta_3 \left( \Gamma_0^A \right)^2 - 8\,\Gamma_0^A\,f(R) \,,
\end{equation}
or equivalently
\begin{equation}\label{notCasi}
   \frac{\dv}{C_A} 
   = \left( \frac{808}{27} + 4\zeta_3 \right) C_A - \frac{224}{27}\,T_F n_f - 32 f(R) \,,
\end{equation}
where $f(R)$ vanishes for $R\to\infty$. For realistic values $R<1$, this function can be very well approximated by the expansion \cite{Banfi:2012yh}
\begin{equation}
\begin{aligned}
   f(R) &= - \left( 1.0963\,C_A + 0.1768\,T_F n_f \right) \ln R
    + \left( 0.6072\,C_A - 0.0308\,T_F n_f \right) \\
   &\hspace{4.7mm}\mbox{}- \left( 0.5585\,C_A - 0.0221\,T_F n_f \right) R^2
    + \left( 0.0399\,C_A - 0.0004\,T_F n_f \right) R^4 + \dots \,.
\end{aligned}
\end{equation} 

It would be desirable to check expression (\ref{d2v}) with an explicit calculation of the coefficient $\dv$ in SCET. This is left for future work. We have pointed out in \cite{Becher:2010tm} that the coefficient $d_2^g$ entering in the small-$q_T$ resummation for Higgs-boson production is related to the corresponding coefficient $d_2^q$ for Drell-Yan production by the Casimir-scaling relation $d_2^q/C_F=d_2^g/C_A$, and that in this ratio only maximally non-abelian color structures arise. Such a scaling relation no longer holds in the presence of a jet veto, because the function $f(R)$ does not share this property. In particular, in the Drell-Yan case $f(R)$ contains some ``abelian'' terms proportional to $C_F$ \cite{Banfi:2012yh}, which give rise to a two-loop $C_F^2$ term in the exponent of the anomaly factor, indicating that this exponent is no longer constrained by the non-abelian exponentiation theorem.

The explicit results collected in this section, along with the known NLO expressions in RG-improved perturbation theory for the hard matching coefficients $C_t$ in (\ref{CtNLO}) and $C_S$ in (\ref{CSevolsol}) and (\ref{eq:CS}), provide the basis for the first NNLL resummation of the Higgs-boson production cross section with a jet veto. In the context of effective field-theory calculations it is conventional to count orders in RG-improved perturbation theory, while the literature on collider physics usually quotes the logarithmic accuracy of the resummation. A resummation with NNLL accuracy in the exponentiated cross section corresponds to a cross-section calculation at NLO in RG-improved perturbation theory. At this order, we need the two-loop anomaly function in (\ref{Fgg2}) but only one-loop expressions for the kernel functions in (\ref{Ires}). Before presenting the numerical results of such an analysis, it is interesting to compare our approach with that recently put forward in \cite{Banfi:2012yh}, where the first resummation of the jet-veto cross section at NLL order was accomplished. The results of these authors can be recovered from our expressions by using the lowest-order approximations
\begin{equation}
   F_{gg}(\pTveto,\mu) = a_s\,\Gamma_0^A\,L_\perp \,, \qquad
   I_{g\leftarrow i}(z,\pTveto,\mu) = \delta(1-z)\,\delta_{gi} \,,
\end{equation}
along with the leading-order RG-improved expressions for the Wilson coefficients $C_t$ (which equals 1 at this order) and $C_S$. Our results go significantly beyond this approximation, and they can in principle be extended to higher orders in a systematic way. We will comment below on the numerical differences between NLL and NNLL calculations for Higgs production at the LHC.

While a two-loop calculation of the kernel functions $I_{g\leftarrow i}$ is beyond the scope of this work, we believe that such a calculation may still be practical using presently available tools. Combined with the known expressions for the hard matching coefficients $C_t$ and $C_S$ at NNLO in RG-improved perturbation theory, this would provide the most important ingredients for a calculation of the jet-veto cross section at N$^3$LL order. (For a complete result, also the four-loop cusp anomalous dimension and the three-loop anomaly coefficient $d_3^{\rm veto}$ would be needed, but their numerical effect is expected to be rather small.) Sensitivity to the jet algorithm arises first in two-loop diagrams in which two intermediate lines are cut, corresponding to final states $X$ containing two collinear, anti-collinear, or soft partons. In the phase-space integrals for these partons, we can change variables such that
\begin{equation}\label{ps}
   \frac{d^3k_i}{(2\pi)^3\,2E_{k_i}}\to \frac{dy_i\,d\phi_i\,k_{Ti}\,dk_{Ti}}{16\pi^3} \,; \quad
   i=1,2 \,,
\end{equation}
where $k_{Ti}=|k_{\perp i}|$, and without loss of generality we can set $\phi_1=0$ and $\phi_2=\phi$. If for the two emitted partons $\sqrt{(y_1-y_2)^2+\phi^2}<R$, then these two partons are merged into a single jet, and the jet veto must be applied to the sum of their transverse momenta, i.e., we must impose that $\sqrt{k_{T1}^2+k_{T2}^2+2k_{T1}k_{T2}\cos\phi}<\pTveto$. This happens inside a disk with radius $R$ in the $y$\,--\,$\phi$ plane. If the first condition is not satisfied, then the partons are classified as two individual jets and the jet veto must be applied to each of them, i.e., we must require that $k_{T1},k_{T2}<\pTveto$. The integrals over transverse momenta can be evaluated after the jet clustering has been performed.

\section{Higgs production at the LHC}
\label{sec:pheno}

We now perform a phenomenological analysis of the jet-veto cross section at the LHC with $\sqrt{s}=8$\,TeV. For concreteness, we consider the cross section integrated over rapidity, given in (\ref{sigfinal}). We set $R=0.4$ and recall that to the order we are working the results are the same for all inclusive, $k_T$-style jet algorithms ($k_T$, anti-$k_T$, or Cambridge/Aachen). The $R$ dependence only enters via the coefficient $d_2^{\rm veto}$ in (\ref{d2v}) and the matching corrections, both of which are NNLO terms which involve at most two real emissions. It follows that the clustering is the same for all version of the jet algorithm. All our numerical predictions are obtained using MSTW2008NNLO PDFs with $\alpha_s(M_Z)=0.1171$ \cite{Martin:2009iq}. While one could argue that it is more natural to evaluate the lower-order predictions with NLO PDF sets, we prefer to keep the PDFs fixed in order to asses the size of higher-order corrections. Furthermore, even the resummed predictions at NLL  order contain the dominant higher-order corrections, so that using NNLO PDF sets seems more appropriate. 

We take into account finite quark-mass effects in the Born-level cross section and the leading electroweak corrections by replacing
\begin{equation}\label{Bornpr}
   \sigma_0(\mu) \to \sigma_0(\mu)\,\kappa_{\rm EW}\,
    \Big| \sum_{q=t,b}\,A(x_q) \Big|^2 \,; \qquad
    x_q = \frac{4m_q^2}{m_H^2} - i\epsilon
\end{equation}
in (\ref{sigfinal}), where
\begin{equation}
   A(x) = \frac32 \left[ x + x(1-x) \arctan^2\frac{1}{\sqrt{x-1}} \right] .
\end{equation}
The value of this function is very close to 1 for the top quark, while $A(x_b)={\cal O}(m_b^2/m_H^2)$ is rather small. We use $m_t=172.6$\,GeV and set the Higgs mass to $m_H=125$\,GeV. For the $b$-quark mass, we use $m_b(m_b)=4.2$\,GeV, which translates to $m_b(m_H)=2.83$\,GeV used when evaluating the cross section. When this is done, the absolute value of the sum $|A(x_t)+A(x_b)|=0.9997$, indicating that quark-mass effects are tiny at Born level. We include electroweak effects as calculated in \cite{Aglietti:2004nj,Degrassi:2004mx,Actis:2008ug} and implemented in the code {\sc RGHiggs} \cite{Ahrens:2010rs} for the total cross section. For $m_H=125$\,GeV, this gives a factor $\kappa_{\rm EW}=1.0514$. Overall, the change in the Born-level cross section obtained from (\ref{Bornpr}) amounts to an enhancement by a factor 1.0507.

\begin{figure}
\begin{center}
\includegraphics[width=0.9\textwidth]{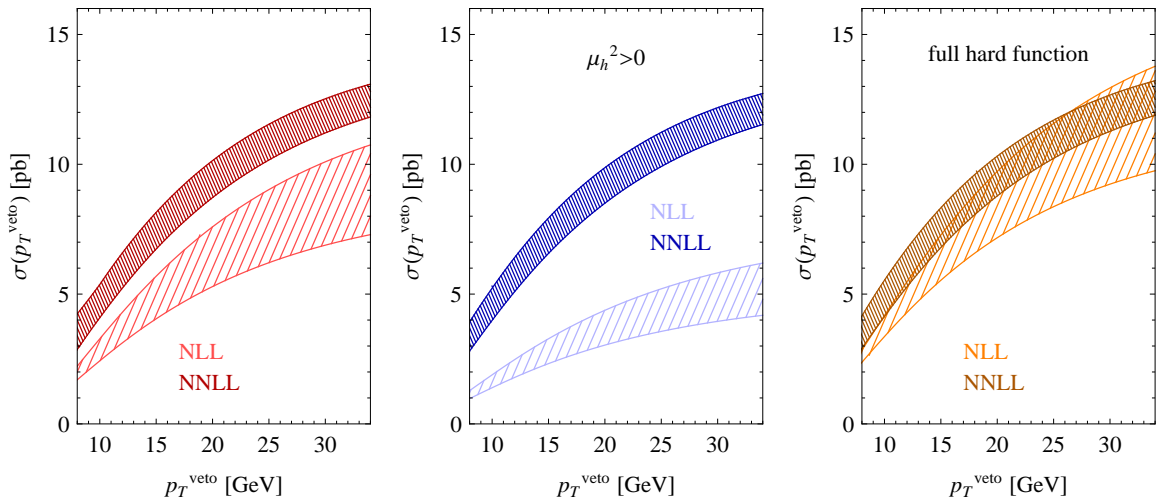}
\caption{\label{fig:resSchemes} 
Resummed predictions for the Higgs-boson production cross section with a jet veto at the LHC (with $\sqrt{s}=8$\,TeV). The plot on the left shows our results at NLL and NNLL order obtained with the default choices of matching scales and scheme. The second plot illustrates the presence of the large corrections encountered when a space-like value of the hard matching scale $\mu_h$ is used. The third plot shows the effect of factoring out the hard function at NNLO.}
\end{center}
\end{figure}

The first plot in Figure~\ref{fig:resSchemes} shows our resummed predictions for the jet-veto cross section at NLL and NNLL order. These results are obtained by evaluating relation (\ref{sigfinal}) using (\ref{Fgg2}) and (\ref{Ires}) as well as the NLO expressions for the hard matching coefficients $C_t$ and $C_S$ obtained in RG-improved perturbation theory. For the default value of the factorization scale we adopt the canonical choice $\mu=\pTveto$, while the default hard matching scales are $\mu_t=m_t$ in (\ref{CtNLO}) and $\mu_h^2=-m_H^2$ in (\ref{CSevolsol}). The bands in the figure are obtained by varying the three scales independently by a factor~2 about their default values and then adding the resulting uncertainties in quadrature. The main effect is due to the variation of the factorization scale $\mu$. At NLO, the change in the cross section from the $\mu_h$ and $\mu_t$ variations is only about 1\%.

Next, we turn to the treatment of the hard function $C_t^2(m_t^2,\mu)\,|C_S(-m_H^2,\mu)|^2$ in the factorization formula (\ref{sigfinal}), which arises as an overall factor both in the jet-veto cross section and the total cross section for Higgs production. As shown in \cite{Ahrens:2008qu,Ahrens:2008nc}, the Wilson coefficient $C_S$ receives very large perturbative corrections, which arise in the analytic continuation of the scalar form factor $C_S(Q^2,\mu_h)$ to the time-like momentum transfer $Q^2=-m_H^2$ relevant for Higgs production. These corrections can be resummed by choosing the time-like value $\mu_h^2=-m_H^2$ for the hard matching scale and then using RG evolution to evolve back to space-like scale choices. The positive effect of this scale choice can be seen by comparing the first and second plots shown in Figure~\ref{fig:resSchemes}. The red bands in the left panel correspond to scale variations about our default choice $\mu_h^2=-m_H^2$, while the blue bands in the middle panel correspond to variations about $\mu_h^2=m_H^2$. With such a choice, the resummed result suffers from the same large higher-order corrections that affect the fixed-order expansion of the total cross section. 

Since the Wilson coefficients $C_t$ and $C_S$ are known even at NNLO in RG-improved perturbation theory, and since they appear as prefactors in the expression for the jet-veto cross section, it is not unreasonable to use the most accurate expressions for these coefficients that is available. Numerically, the difference between taking the NLO and NNLO hard functions is not very large. Including the NNLO corrections leads to a 2\% increase of the cross section at low $\pTveto$ and a 1\% increase at higher values. Using the NNLO approximation for the hard function will also be convenient below, when we match our resummed expression to fixed-order perturbation theory. From here on, we write the RG-improved hard function in the exponentiated form
\begin{equation}\label{logH}
   C_t^2(m_t^2,\mu)\,|C_S(-m_H^2,\mu)|^2
   = \exp\Big[ 2g_1(m_t^2,\mu_t,\mu) + 2\mbox{Re}\,g_2(-m_H^2,\mu_h,\mu) \Big] \,,
\end{equation}
where the functions $g_1$ and $g_2$ will be expanded to NNLO in the exponent. The corresponding expressions are given in the appendix. The last plot in Figure~\ref{fig:resSchemes} shows that this treatment improves the convergence behavior of the expansion for the resummed cross section, since the bands obtained at NLL and NNLL order now overlap. We stress, however, that while the scheme ambiguities reflected by the results in the figure are significant when the resummation is performed at NLL order, the differences between the various schemes become negligible once we perform the resummation at NNLL order. This demonstrates the importance of our analysis, which for the first time has enabled a resummation with this accuracy. 

\begin{figure}
\begin{center}
\includegraphics[width=0.9\textwidth]{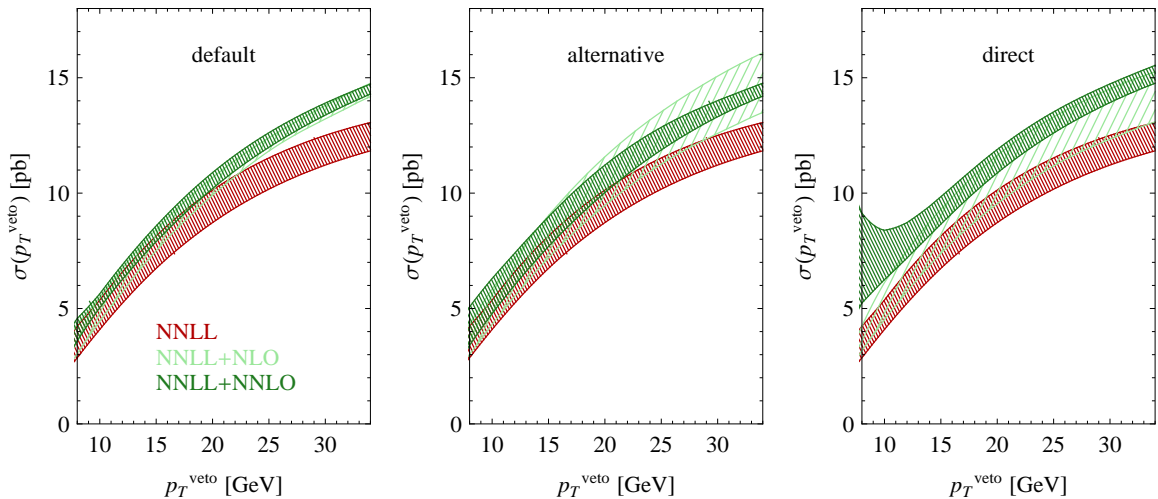}
\caption{\label{fig:schemes} 
Cross-section predictions obtained in the default matching scheme (left) and an alternative scheme (center), in which also the anomaly term is factored out before the matching to fixed-order perturbation theory is performed. The right plot illustrates the bad perturbative behavior encountered when the hard function is not factored out.}
\end{center}
\end{figure}

To obtain the best possible predictions for the jet-veto cross section, we  match our resummed expression (\ref{sigfinal}) to the known NNLO fixed-order result for the cross section, which can be obtained by running the codes {\sc FeHiP} \cite{Anastasiou:2005qj} or {\sc HNNLO} \cite{Catani:2007vq,Grazzini:2008tf}. In our numerical analysis below we use the latter. The matching to fixed order allows us to also include terms that are suppressed by powers of $\pTveto/m_H$, which are not captured by our leading-power factorization formula. In addition to the power-suppressed terms, the matching also provides us with the two-loop constant, which is not predicted by a resummation at NNLL order. To ensure that the matching correction is not contaminated by the large perturbative corrections contained in the hard function, we will factor out the hard function (as discussed above) and define a reduced cross section $\widetilde\sigma(\pTveto,\mu)$ as
\begin{equation}\label{sigRed}
   \sigma(\pTveto) = C_t^2(m_t^2,\mu) \left| C_S(-m_H^2,\mu) \right|^2
    \times \widetilde\sigma(\pTveto,\mu) \,.
\end{equation}
To remove the prefactor from the fixed-order result, we need to divide it by the two-loop fixed-order expansion of the hard function, which is given in (\ref{hardfix}) in the  appendix. The matching is then performed for the reduced cross section according to 
\begin{equation}
   \widetilde\sigma_{\rm NNLL+NNLO} 
   = \widetilde\sigma_{\rm NNLL} + \left( \widetilde\sigma_{\rm NNLO} 
    - \widetilde\sigma_{\rm NNLL}\big|_{\text{expanded to NNLO}} \right) .
\end{equation}
Our result for the resummed cross section (at NNLL order) matched to fixed-order perturbation theory (at NLO and NNLO) is shown in the first plot in Figure~\ref{fig:schemes}. We observe that the matching corrections become more significant at larger values of the jet veto. This is expected, since the power corrections are controlled by the ratio $\pTveto/m_H$. Once the resummation is implemented in the way discussed above, the dominant matching corrections arise at NLO. The additional effect of performing the matching at two-loop order is to further reduce the residual scale dependence slightly. The second plot in the figure shows our results in a different scheme, in which in addition to the hard function also the anomaly term -- the last factor in the first line of (\ref{sigfinal}) --  is factored out before the matching to fixed-order perturbation theory is performed. Comparing the two plots, we observe that the results are very similar, even though the scale variation at NNLL+NLO is somewhat larger in the second case. This is to be contrasted with the results shown in the last panel in Figure~\ref{fig:schemes}, which are obtained if one neglects to factor out the hard function in the way shown in (\ref{sigRed}). In this case, the matching corrections suffer from a rather bad perturbative behavior, which becomes particularly obvious at small values of $\pTveto$. 
 
\begin{figure}
\begin{center}
\includegraphics[width=0.9\textwidth]{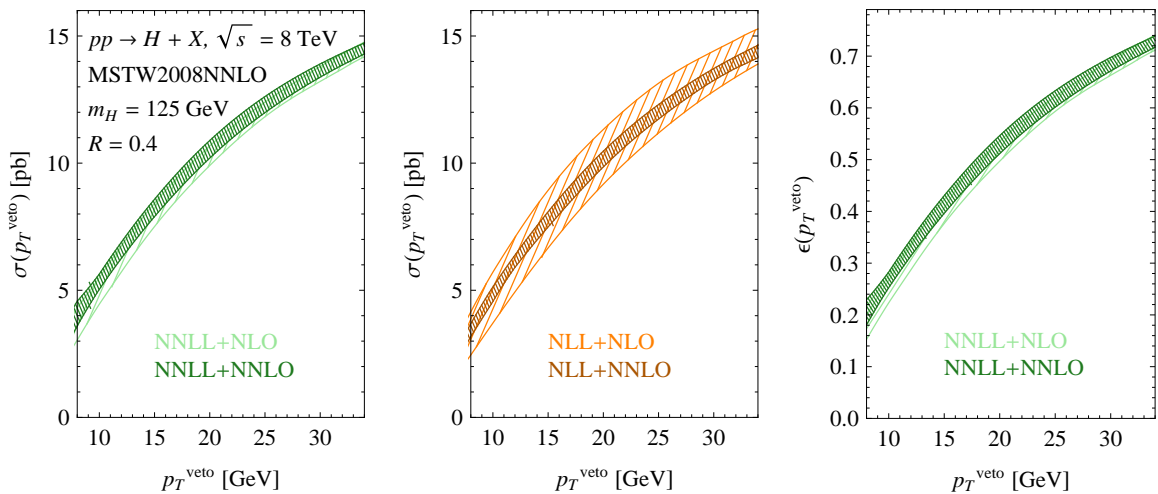}
\caption{\label{fig:matching} 
Resummed and matched predictions for the jet-veto cross section (first two plots) and efficiency (last plot) for Higgs-boson production at the LHC.}
\end{center}
\end{figure}

The first plot in Figure~\ref{fig:matching} shows our final results for the resummed and matched jet-veto cross section. For comparison, in the second panel we also give the result when the resummation is only performed at NLL order. After matching to NNLO, the difference between the two results is rather small but not entirely negligible. For example, at $\pTveto=20$\,GeV the band obtained at NNLL+NNLO is approximately 4\% higher than at NLL+NNLO. However, at NNLL order the matching corrections are much smaller. The excellent stability of the result is comforting, but is in part due to our improved matching scheme, in which the hard function is always taken at NNLO in RG-improved perturbation theory. Then the difference between the matched NLL and NNLL results is only due to higher-order terms from the anomaly. In Table~\ref{tab:sigmanum}, we give predictions for the jet-veto cross section including estimates of the perturbative uncertainty (shown by the bands in Figure~\ref{fig:matching}) as well as PDF and $\alpha_s$ uncertainties. The latter are estimated by re-evaluating the cross section with a number of error PDFs, which were fitted to data for different $\alpha_s$ values and encode the experimental uncertainties in the data sets entering the PDF fits \cite{Martin:2009iq}. The uncertainty on the cross section is then computed from these results according to the prescription associated with a given PDF set. We find that the relative PDF and $\alpha_s$ uncertainties are rather insensitive to higher-order effects. To excellent accuracy, they can be inferred from the Born-level cross section evaluated at the low scale $\mu=\pTveto$. It is thus a simple matter to compute the PDF uncertainty also for other PDF sets.

We emphasize that in our approach a prediction is obtained directly for the Higgs-boson production cross section with a jet veto -- the observable that hopefully will soon be measured  at the LHC. In contrast, the authors of \cite{Banfi:2012yh} focused on the jet-veto efficiency, defined as $\epsilon(\pTveto)=\sigma(\pTveto)/\sigma_{\rm tot}$. In order to compare to their results, we now briefly discuss the veto efficiency as well. For the total cross section, we use the central value of the result obtained in \cite{Ahrens:2010rs} and implemented in the code {\sc RGHiggs} \cite{Ahrens:2010rs}. At $\sqrt{s}=8$\,TeV and $m_H=125$\,GeV, we obtain $\sigma_{\rm tot}=(19.65^{+0.55+1.54}_{-0.15-1.48})$\,pb, where the second error gives the combined PDF and $\alpha_s$ uncertainties at 90\% CL. The resummed expressions for the jet-veto and the total cross sections are quite similar. They involve exactly the same hard function, and to NLL order the relative PDF and $\alpha_s$ uncertainties are the same when evaluated at the same factorization scale. These uncertainties therefore cancel to a large extent when taking the ratio of $\sigma(\pTveto)$ and $\sigma_{\rm tot}$, and hence they can be neglected in the predictions for the jet-veto efficiency. The scale uncertainties in the two cross sections are correlated in such a way that the relative error on $\epsilon(\pTveto)$ would be slightly smaller than that on $\sigma(\pTveto)$. To be conservative, we normalize by the central value of the total cross section and retain the full scale uncertainty of the jet-veto cross section. The corresponding results are shown in the third plot in Figure~\ref{fig:matching} as well as in the last column of Table~\ref{tab:sigmanum}. To a good approximation, one can infer the jet-veto cross sections corresponding to other PDF sets by multiplying the veto efficiency with the central value for the resummed total cross section and add its PDF uncertainties. The corresponding values for $\sigma_{\rm tot}$ can be obtained by running the {\sc RGHiggs} code with a different set of PDFs. For example, we find $\sigma_{\rm tot}=(20.87_{ -0.35-1.43}^{+0.63+1.43})$\,pb using NNPDF2.0 \cite{Ball:2010de}, and $\sigma_{\rm tot}=(19.72_{-0.15-1.01}^{+0.56+0.92 })$\,pb using CTEQ6.6 PDFs \cite{Lai:2010nw}. 

\begin{table}
\begin{center}
\renewcommand{\arraystretch}{1.5}
\begin{tabular}{lll}
$\pTveto$ & $\phantom{a}\sigma(\pTveto)$ [pb] & $\phantom{a}\epsilon(\pTveto)$  \\\hline
$10$ & $\phantom{0}4.97_{-0.01-0.22}^{+0.59+0.19}$ & $0.253_{-0.001}^{+0.030}$ \\ 
$15$ & $\phantom{0}7.83_{-0.05-0.35}^{+0.62+0.34}$ & $0.399_{-0.002}^{+0.032}$ \\ 
$20$ & $10.13_{-0.08-0.49}^{+0.61+0.48}$ & $0.515_{-0.004}^{+0.031}$ \\ 
$25$ & $11.88_{-0.07-0.61}^{+0.55+0.61}$ & $0.605_{-0.003}^{+0.028}$ \\ 
$30$ & $13.23_{-0.03-0.70}^{+0.48+0.71}$ & $0.673_{-0.002}^{+0.024}$ \\
$\infty$ & $19.66_{ -0.15-1.48}^{+0.55+1.54}$ & \hspace{6mm} 1 \\  
\end{tabular}
\end{center}
\caption{\label{tab:sigmanum} 
Predictions for the NNLL+NNLO resummed and matched cross section and efficiency. The second error on the cross section gives the combined PDF and $\alpha_s$ uncertainties at 90\%~CL.}
\end{table}

Our perturbative results for the jet-veto cross section have a high accuracy: the scale uncertainty from varying the factorization scale $\mu$ by a factor 2 around $\mu=\pTveto$ is about 12\% for $\pTveto=10$\,GeV and decreases to 4\% at $\pTveto=30$\,GeV. This is almost as small as the scale uncertainty on the total cross section, which is 3\%. The uncertainty from varying the hard matching scales $\mu_t$ and $\mu_h$ is negligible, since we evaluate the hard function at NNLO in RG-improved perturbation theory. Indeed, by now also the three-loop correction to $C_S(-m_H^2-i\epsilon,\mu_h)$ is known \cite{Baikov:2009bg,Lee:2010cga,Gehrmann:2010tu}, and it amounts to a mere $-0.015\%$ shift for $\mu_h^2=-m_H^2$. (With $\mu_h^2=+m_H^2$, the correction would be $4\%$, which confirms once again the advantage of using a time-like scale choice.) While the logarithmic accuracy of the resummation is one order higher for the total cross section, our result for the jet-veto cross section includes the most crucial part, the hard function, at the same accuracy, and so it should not come as a surprise that the scale uncertainties turn out to be comparable. 

Finally, let us relate our findings to the NLL+NNLO results presented in the recent paper \cite{Banfi:2012yh}. An important difference to our work is that these authors do not compute the cross section directly, but instead present results for the jet-veto efficiency, arguing that its uncertainties are largely uncorrelated with those of the total cross section. In that way they try to avoid the large perturbative corrections known to affect the $gg\to H$ amplitude. Indeed, since the hard function is the same for the jet-veto and the total cross section, the corresponding large corrections formally cancel in the ratio of these two quantities. However, the extent to which this cancellation takes place depends on the precise way in which the perturbative expansion of the efficiency is performed. Equivalent matching schemes therefore lead to sizable differences in their results. Specifically, the authors use three matching schemes, related to different definitions of the fixed-order expansion of the efficiency $\epsilon(\pTveto)$. In their scheme ($a$), the efficiency is obtained by expanding both the resummed jet-veto cross section and the fixed-order total cross section to a given order and then taking the ratio of the two. In scheme ($c$), the perturbative expansion is performed directly for the efficiency, while scheme ($b$)  infers the efficiency from the normalized 1-jet rate. The fixed-order result for ($a$) and ($c$) is shown in the right two panels of Figure~\ref{fig:resFixed}. As we discussed in the introduction, the difference between these two  schemes is so large because the fixed-order result for the jet-veto cross section involves two types of large corrections: those from the analytic continuation of the hard function, which also affect the total cross section, and those from Sudakov logarithms associated with the jet veto. Different schemes combine these in different ways, which then leads to quite different matching corrections. Even for $\pTveto=20$\,GeV, the difference between schemes ($b$) and ($c$) is more than 20\%, which is shocking given that it is formally an ${\cal O}(\alpha_s^3)$ effect. The origin of this problem is that the authors of \cite{Banfi:2012yh} resum the Sudakov logarithms associated with the jet veto, but they do not control the large corrections to the total cross section.

\section{Summary}

We have presented for the first time a factorization theorem for the Higgs-boson production cross section in the presence of a veto against jets with transverse momentum above a threshold $\pTveto$. It forms the basis for the systematic resummation of Sudakov logarithms $\alpha_s^n\ln^k(m_H/\pTveto)$ with $k\le 2n$ beyond the LL approximation. In analogy with the situation encountered in the analysis of the transverse-momentum spectrum of the Higgs boson, the final expressions (\ref{dsigfinal}) and (\ref{sigfinal}) are affected by a collinear anomaly, which introduces an extra power-like dependence on the hard scale $m_H$. We have derived the all-order form of this anomaly and have determined its two-loop expression by combining information on the two-loop transverse-momentum spectrum of the Higgs boson with a recent result for certain jet-radius dependent two-loop terms obtained by Banfi, Salam and Zanderighi in \cite{Banfi:2012yh}. Using results on the RG-improved hard matching functions available from previous studies of Higgs production in SCET, we have been able to perform the first resummation of the jet-veto cross section and efficiency at NNLL accuracy. These results have then been matched to the available NNLO fixed-order result for the cross section, which accounts for power corrections in the ratio $\pTveto/m_H$ as well as the non-logarithmic two-loop contributions.

In addition to Sudakov logarithms of the ratio of $m_H/\pTveto$, the cross section also suffers from large perturbative corrections to the hard function, which arise in the analytic continuation of the scalar gluon form factor to the time-like kinematics relevant for the $gg\to H$ amplitude. The same corrections also affect the total cross section. They can be resummed in a systematic way using RG methods \cite{Ahrens:2008qu,Ahrens:2008nc}. To obtain the most accurate prediction possible, we have performed the matching of our NNLL result to fixed-order perturbation theory after the hard function has been factored out, as shown in (\ref{sigRed}). Once this is done, the matching corrections are quite small, and we observe a very good perturbative stability of our results.

Our phenomenological results presented in Section~\ref{sec:pheno} demonstrate that, with our improved formalism, we are in a position to predict the jet-veto cross section with high accuracy. We stress that we obtain the physically measurable cross section directly, and not by combining a prediction for the veto efficiency with a separate calculation of the total cross section. In fact, at least in one respect the theoretical analysis of the jet-veto cross section is simpler than that of the total cross section: it is evident that resummation is necessary and that non-logarithmic terms are strongly suppressed at small $\pTveto$. The resummation of the total cross section, on the other hand, is based on an expansion around the partonic threshold, and it is not {\it a priori\/} clear how much of the cross section is generated by this region. Our final results for the cross section with a jet veto, shown in Figure~\ref{fig:matching} and collected in Table~\ref{tab:sigmanum}, exhibit scale uncertainties decreasing from 12\% at $\pTveto=10$\,GeV to 4\% at $\pTveto=30$\,GeV. Additional uncertainties from PDFs and the value of $\alpha_s$ are of a similar magnitude. The uncertainty we achieve is similar to the one for the total cross section.

The results obtained in this paper form the basis for model-independent, high-precision calculations of the Higgs-boson production cross section in the presence of a jet veto. We believe that, using available theoretical methods, it may be possible to extent the NNLL+NNLO analysis presented here to include the dominant logarithmic corrections arising at N$^3$LL order. Even without such an improvement, however, the accuracy of our predictions is sufficiently high to allow for a reliable prediction of the cross section, which can hopefully soon be confronted with experimental data.

\vspace{4mm}\noindent
{\em Acknowledgements:\/}
We are grateful to Andrea Banfi, Guido Bell, Ben Pecjak, Lorena Rothen, Gavin Salam, Jonathan Walsh, Giulia Zanderighi, and Saba Zuberi for useful discussions. The research of M.N.\ is supported by the Advanced Grant EFT4LHC of the European Research Council (ERC), grant NE 398/3-1 of the German Research Foundation (DFG), grant 05H09UME of the German Federal Ministry for Education and Research (BMBF), and the Rhineland-Palatinate Research Center {\em Elementary Forces and Mathematical Foundations}. The work of T.B.\ is supported by the Swiss National Science Foundation (SNF) under grant 200020-140978 and the ÒInnovations- und Kooperationsprojekt C-13Ó of the Schweizerische Universit\"atskonferenz (SUK/CRUS).

\vspace{4mm}\noindent
{\bf Note added in proof:\/}
After this paper was accepted for publication, reference \cite{Banfi:2012jm} appeared, in which a NNLL resummation formula for the jet-veto cross section was derived that is equivalent to our result (\ref{sigfinal}). However, in this work it was also shown that the result (\ref{fromGiulia}), which we have taken from \cite{Banfi:2012yh}, is modified in the limit of large $R$. The additional contribution is such that the term proportional to $\zeta_3$ in (\ref{d2v}) gets cancelled. The correct relation then reads $\dv=d_2^g-8\Gamma_0^A f(R)$, and as a result in (\ref{notCasi}) the coefficient of $\zeta_3$ in the first term becomes $-28$ instead of 4. The numerical impact of this correction on our results will be studied elsewhere.

In the recent paper \cite{Tackmann:2012bt} it was claimed that the clustering of soft and collinear particles into the same jet is unsuppressed for $R={\cal O}(1)$, and hence our factorization formula (\ref{sigfinal}) would break down. If correct, this observation would imply that neither our result nor that of \cite{Banfi:2012jm} correctly reproduce the structure of the cross section at NNLL order. We have convinced ourselves that the soft-collinear mixing terms are absent if the multipole expansion is correctly implemented in SCET, or equivalently, if the double counting of overlapping momentum regions is removed by performing appropriate subtractions. For the example of two independent emissions studied in \cite{Tackmann:2012bt}, we have explicitly rederived the result (4) of \cite{Banfi:2012jm} from a SCET analysis based on our formula (\ref{sigfinal}). We also note that the authors of \cite{Banfi:2012jm} have checked their NNLL corrections against available fixed-order results up to ${\cal O}(\alpha_s^3)$.

\newpage
\begin{appendix}

\section{Results for the hard matching coefficients}
\renewcommand{\theequation}{A\arabic{equation}}
\setcounter{equation}{0}

The Wilson coefficient $C_t$ in the effective Lagrangian (\ref{Leff}) satisfies the RG equation \cite{Inami:1982xt}
\begin{equation}
   \frac{d}{d\ln\mu}\,C_t(m_t^2,\mu^2) = \gamma^t(\alpha_s)\,C_t(m_t^2,\mu^2) \,,
    \qquad \mbox{with} \quad
   \gamma^t(\alpha_s) = \alpha_s^2\,\frac{d}{d\alpha_s}\,\frac{\beta(\alpha_s)}{\alpha_s^2} \,.
\end{equation}
The anomalous dimension is related to the QCD $\beta$-function $\beta(\alpha_s)=d\alpha_s/d\ln\mu$, because the two-gluon operator in the effective Lagrangian is the Yang-Mills Lagrangian. The exact solution to the evolution equation reads
\begin{equation}
   C_t(m_t^2,\mu)
   = \frac{\beta\big(\alpha_s(\mu)\big)/\alpha_s^2(\mu)}%
          {\beta\big(\alpha_s(\mu_t)\big)/\alpha_s^2(\mu_t)}\,
   C_t(m_t^2,\mu_t) \,,      
\end{equation}
where $\mu_t\sim m_t$ is a hard matching scale at which the Wilson coefficient can be evaluated in fixed-order perturbation theory. The two-loop expression for $C_t(m_t^2,\mu_t)$ was calculated in \cite{Kramer:1996iq,Chetyrkin:1997iv} and is given in eq.~(12) of \cite{Ahrens:2008nc}. The three-loop expression is also known \cite{Schroder:2005hy,Chetyrkin:2005ia}.

The RG equation for the SCET matching coefficient of the gluonic current in (\ref{current}) reads~\cite{Ahrens:2008nc}
\begin{equation}
   \frac{d}{d\ln\mu}\,C_S(-m_H^2-i\epsilon,\mu)
   = \left[ \Gamma_{\rm cusp}^A(\alpha_s)\,
    \ln\frac{-m_H^2-i\epsilon}{\mu^2} + \gamma^S(\alpha_s) \right] 
    C_S(-m_H^2-i\epsilon,\mu) \,,
\end{equation}
and its general solution has the form 
\begin{equation}\label{CSevolsol}
   C_S(-m_H^2-i\epsilon,\mu) 
   = \exp\left[ 2S(\mu_h,\mu) - a_\Gamma(\mu_h,\mu)\,\ln\!\frac{-m_H^2-i\epsilon}{\mu_h^2}
    - a_{\gamma^S}(\mu_h,\mu) \right] C_S(-m_H^2-i\epsilon,\mu_h) \,,
\end{equation}
where 
\begin{equation}\label{eq:CS}
   C_S(-m_H^2-i\epsilon,\mu_h) 
   = 1 + \frac{\alpha_s(\mu_h)}{4\pi}\,C_A 
    \left( -\ln^2\frac{-m_H^2-i\epsilon}{\mu_h^2} + \frac{\pi^2}{6} \right) + \dots
\end{equation}
is the matching condition at an appropriately chosen hard scale $\mu_h$, at which the Wilson coefficient has a well-behaved perturbative series. We have demonstrated in \cite{Ahrens:2008nc} that this is not the case when one adopts the naive choice $\mu_h\sim m_H$. On the other hand, a time-like scale choice such that $\mu_h^2\sim-m_H^2$ gives rise to a nicely convergent series. The fact that $\alpha_s(\mu_h)$ is complex in this case does not pose any difficulty for the numerical evaluation. The two-loop corrections to (\ref{eq:CS}) can be derived from the results of \cite{Harlander:2000mg} and have been given in eq.~(17) of \cite{Ahrens:2008nc}. The three-loop corrections are also known \cite{Baikov:2009bg,Lee:2010cga,Gehrmann:2010tu}. In the solution (\ref{CSevolsol}) we have introduced the definitions \cite{Neubert:2004dd}
\begin{equation}
   S(\nu,\mu) 
   = - \int\limits_{\alpha_s(\nu)}^{\alpha_s(\mu)}\!
    d\alpha\,\frac{\Gamma_{\rm cusp}^A(\alpha)}{\beta(\alpha)}
    \int\limits_{\alpha_s(\nu)}^\alpha \frac{d\alpha'}{\beta(\alpha')} \,, \qquad
   a_\Gamma(\nu,\mu) 
   = - \int\limits_{\alpha_s(\nu)}^{\alpha_s(\mu)}\!
    d\alpha\,\frac{\Gamma_{\rm cusp}^A(\alpha)}{\beta(\alpha)} \,, 
\end{equation}
and similarly for the function $a_{\gamma^S}$. The perturbative expansions of these functions obtained at NNLO in RG-improved perturbation theory can be found in eqs.~(93) and (94) in the appendix of \cite{Becher:2006mr}. The required anomalous-dimension coefficients entering these expressions are given, e.g., in eqs.~(A2) of \cite{Ahrens:2008nc} and (95) of \cite{Becher:2006mr}. Up to three-loop order, the cusp anomalous dimension in the adjoint representation differs from that in the fundamental representation simply by an overall factor $C_A/C_F$.

The functions $g_1$ and $g_2$ entering the exponent in (\ref{logH}) follow from the above results. We find 
\begin{equation}
\begin{aligned}
   g_1(m_t^2,\mu_t,\mu) 
   &= \ln\frac{\beta\big(\alpha_s(\mu)\big)/\alpha_s^2(\mu)}%
              {\beta\big(\alpha_s(\mu_t)\big)/\alpha_s^2(\mu_t)} 
    + \ln C_t(m_t^2,\mu_t) \,, \\
   g_2(-m_H^2,\mu_h,\mu) 
   &= 2S(\mu_h,\mu) - a_\Gamma(\mu_h,\mu)\,\ln\!\frac{-m_H^2}{\mu_h^2}
    - a_{\gamma^S}(\mu_h,\mu) + \ln C_S(-m_H^2,\mu_h) \,.
\end{aligned}
\end{equation}

For the matching of our resummed result for the jet-veto cross section to the corresponding fixed-order expression, it is necessary to derive a fixed-order result for the reduced cross section $\widetilde\sigma(\pTveto,\mu)$ defined in (\ref{sigRed}), in which the hard function is factored out. To compute it to NNLO, we need to divide the fixed-order result by the two-loop expansion of the hard function. For $N_c=3$ colors and $n_f=5$ light quark flavors (the top quark is integrated out), we obtain
\begin{equation}\label{hardfix}
   C_t^2(m_t^2,\mu) \left| C_S(-m_H^2,\mu) \right|^2 
   = 1 + c_1\,\frac{\alpha_s(\mu)}{4\pi} + c_2 \left( \frac{\alpha_s(\mu)}{4\pi} \right)^2 
    + \dots \,,
\end{equation}
where
\begin{equation}
\begin{aligned}
   c_1 &= - 6 L_h^2 + 22 + 7\pi^2  \\
   c_2 &= 18 L_h^4 + \frac{46}{3} L_h^3
    + \left( - \frac{698}{3} - 36\pi^2  \right) L_h^2+ \left( \frac{1240}{9} - \frac{184\pi^2}{3} - 36\zeta_3 \right) L_h \\
   &\quad\mbox{}
    - \frac{274}{3}\,L_t  +\frac{10718}{27}+\frac{1679 \pi ^2}{6}-\frac{998 \zeta_3}{3}+\frac{37 \pi ^4}{2}  \,.
\end{aligned}
\end{equation}
Here $L_h=\ln(m_H^2/\mu^2)$ and $L_t=\ln(m_t^2/\mu^2)$ are the large logarithms. 

\end{appendix}

\end{document}